\documentclass[conference]{IEEEtran}
\usepackage{cite}
\ifCLASSINFOpdf
   \usepackage[pdftex]{graphicx}
\else
\fi

\usepackage{amsmath}
\usepackage{mathtools}
\usepackage{booktabs}
\usepackage[linesnumbered,ruled]{algorithm2e}

\usepackage{stfloats}
	
\usepackage{xcolor}

\usepackage{url}

\begin{document}
%
\title{Customer Lifetime Value in Video Games Using Deep Learning and Parametric Models}

\IEEEoverridecommandlockouts

\author{\IEEEauthorblockN{Pei Pei Chen\IEEEauthorrefmark{1}\IEEEauthorrefmark{3}, Anna Guitart\IEEEauthorrefmark{1}\IEEEauthorrefmark{3}\thanks{\IEEEauthorrefmark{3}These two authors contributed equally to the work.}, Ana Fern\'andez del R\'io\IEEEauthorrefmark{1}\IEEEauthorrefmark{2} and  \'{A}frica Peri\'a\~{n}ez\IEEEauthorrefmark{1}}
\IEEEauthorblockA{\IEEEauthorrefmark{1}%
  Yokozuna Data, a Keywords Studio\\
  102-0074 Tokyo, Chiyoda 5F, Aoba No. 1 Bldg., Japan
 \IEEEauthorblockA{\IEEEauthorrefmark{2}%
  Dpto. F\'isica Fundamental, UNED, Madrid, Spain}
  Email: \{ppchen, aguitart, afdelrio and aperianez\}@yokozunadata.com}
}

\maketitle

\begin{abstract}
Nowadays, video game developers record every virtual action performed by their players. As each player can remain in the game for years, this results in an exceptionally rich dataset that can be used to understand and predict player behavior. In particular, this information may serve to identify the most valuable players and foresee the amount of money they will spend in in-app purchases during their lifetime. This is crucial in free-to-play games, where up to 50\% of the revenue is generated by just around 2\% of the players, the so-called whales. 

To address this challenge, we explore how deep neural networks can be used to predict customer lifetime value in video games, and compare their performance to parametric models such as Pareto/NBD. Our results suggest that convolutional neural network structures are the most efficient in predicting the economic value of individual players. They not only perform better in terms of accuracy, but also scale to big data and significantly reduce computational time, as they can work directly with raw sequential data and thus do not require any feature engineering process. This becomes important when datasets are very large, as is often the case with video game logs.

Moreover, convolutional neural networks are particularly well suited to identify potential whales. Such an early identification is of paramount importance for business purposes, as it would allow developers to implement in-game actions aimed at retaining big spenders and maximizing their lifetime, which would ultimately translate into increased revenue.

\end{abstract}


%
\IEEEpeerreviewmaketitle

\section{Introduction}
Lifetime value (LTV), also called customer lifetime value or lifetime customer value, is an estimate first introduced in the context of marketing \cite{shaw1988database,dwyer1997customer,berger1998customer,hoekstra1999lifetime}, used to determine the expected revenue customers will generate over their entire relationship with a service \cite{pfeifer2005customer}. LTV has been used in a variety of fields---including video games---and is a useful measure for deciding on future investment, personalized player retention strategies, and marketing and promotion plans \cite{farris2010marketing}, as it helps marketers identify potential high-value users.


Models for LTV can be divided into historical and predictive approaches. The former calculate the value of a user based only on their past purchases, without addressing future behavioral changes. On the contrary, predictive schemes do consider potential future variations in the behavior of users to try to predict their purchasing dynamics, taking also into account their lifetime expectancy.

The fundamental elements in historical LTV computations originally come from \emph{RFM models} \cite{fader2005rfm}, which group customers based on \emph{recency}, \emph{frequency}, and \emph{monetary value}---namely, on how recently and how often they purchased and how much they spent. The basic assumption of RFM models is that users with more recent purchases, who purchase more often or who spend larger amounts of money are more likely to purchase again in the future.

Probabilistic models for predicting LTV assume a repeated purchase pattern until the end of the business relationship (i.e.\ until the player \emph{churns}---leaves the game---in our case), and for this reason they are also known as ``buy till you die'' (BTYD)  models \cite{schmittlein1987counting}. One of the most popular formulations is the Pareto/NBD model, which applies to non-contractual and continuous (the customer can purchase at any time) settings, like video games. This method combines two different parametric distributions: the Pareto distribution to obtain a binary classification (indicating whether the customer is still active or not, the so-called dropout process) and the negative binomial distribution (NBD) to estimate the purchase frequency \cite{fader2005note}. 

Other parametric models use different probability distributions---which can be better suited for some problems---to model the dropout (churn) and transaction rates, always operating under the same RFM philosophy. These may involve simplifications that make computations more efficient without giving up significant predictive power---as in beta-geometric (BG) formulations \cite{gupta1991stochastic,platzer2016ticking}---as well as extensions that take into account more details of the customer's transaction history, such as purchase
regularity. An example of the latter is the use of condensed negative binomial distributions (CNBD) \cite{chatfield1973consumer} to model the number of expected transactions. 

Extended Pareto/NBD methods that contain RFM information include a submodel for the amount spent per transaction, besides trying to estimate the number of future transactions per user by fitting probabilistic models to predict their purchasing behavior. 
However, the most common approaches involve simpler computations, such as deriving LTV by cohorts \cite{kumar2004customer} or applying logistic regression with RFM  \cite{mccarty2007segmentation}.

More recent works on LTV prediction make use of machine learning models. For example, \cite{chamberlain2017customer} employed random forests to estimate the LTV of the costumers of an online fashion retailer, while \cite{tkachenko2015autonomous} applied deep Q-learning to predict lifetime donations from clients who received donation-seeking mailings.\looseness=-1

In the last few years, free-to-play (F2P) games that can be downloaded and played for free have become one of the major business models in the video game industry. In these games, most of the revenue is generated through in-app purchases, as users can buy items or other game-related advantages (e.g. playing ad-free). This is precisely the kind of freemium setup to which parametric models apply. 

LTV calculations are also customary in the context of video games. For instance, in \cite{luton2013free,davidovici2013innovation}, a formulation to derive LTV from the average revenue per user and the expected lifetime is presented. However, there are only a few studies using machine learning models in this field. The authors of \cite{sifa2015predicting} use a binary classifier to predict whether purchases will happen, and then a regression model to estimate the number of future purchases made by players. Also, in \cite{voigt2016making}, machine learning methods are combined with the synthetic minority oversampling technique (SMOTE) to predict LTV for individual players. The LTV of mobile game players was calculated through gradient boosting in \cite{drachen2018or}. Finally, in a recent work \cite{sifa2018customer}, Sifa et al. evaluate different machine learning methods together with SMOTE to obtain accurate 
predictions of a user's cumulative spend during one year. 

\subsection{Our contribution}

This work assesses the potential of deep learning to predict player LTV in production settings, and the added value it can bring through the early detection of high-value players. We use a deep perceptron multilayer network and convolutional neural networks (CNNs) to predict the in-app purchases of players based on their playing behavioral records, and then estimate their economic value over the next year. The results are compared to several parametric models, including Pareto/NBD and some popular extensions of it, such as the BG/CNBD (beta-geometric/condensed negative binomial distribution) or MBG/CNBD (where the ``M'' stands for ``modified'') models. To the best of our knowledge, this is the first work using CNNs to predict LTV in the context of video games (Sifa et~al.\ presented an LTV study using deep perceptron multilayer networks \cite{sifa2018customer}, but they did not employ CNNs) and also the first one that carries out an extensive comparison between deep learning techniques and the parametric models traditionally used to calculate LTV in other fields.
In this article we show how CNNs emerge as the most suitable method to predict LTV in terms of accuracy.

\section{Model Description}

\subsection{Pareto/NBD model}

The Pareto/NBD model \cite{schmittlein1987counting} is the most popular BTYD formulation. This parametric approach is designed to predict the number of purchases a customer will make up to a certain time based on their purchase history. Transaction and dropout rates are assumed to vary independently across customers, and this heterogeneity is modeled by means of gamma distributions \cite{wheat1990estimating}. The shape and scale parameters of these two distributions are the four parameters that will be used, together with the customer transaction history, to predict future purchase behavior \cite{schmittlein1987counting,fader2005note}. By using a gamma--gamma submodel for the spend per transaction, the LTV value can be finally obtained \cite{fader2005rfm}.    

The transaction history of each customer is described using only two quantities: the number of purchases made up to that moment (frequency) and the time of their last purchase (recency). Additionally, it is necessary to specify the total time each customer has been observed, i.e.\ the time since their first purchase, as an input. Assuming a Poisson distribution for the number of purchases at a given time, with each customer having their own transaction rate, their continuous mixture gives rise to a negative binomial distribution (NBD). Similarly, considering that the expected lifetime is exponentially distributed, with each customer having their own dropout rate, makes the probability of being active at a certain time follow a Pareto distribution. The maximum likelihood function of the model can be then derived and the four relevant parameters estimated through its maximization. The number of future purchases can be then predicted for every customer as conditional expectations on the estimated parameters and frequency, recency and observed time for each particular customer. Namely, as the conditioned expected number of purchases times the conditioned probability of the particular customer being active.



\subsection{Other parametric models}
The BG/CNBD \cite{platzer2016ticking} and MBG/CNBD \cite{batislam2007empirical} models are extensions of the Pareto/NBD model. The former, which combines the beta-geometric \cite{gupta1991stochastic,platzer2016ticking} and condensed negative binomial distributions \cite{chatfield1973consumer}, increases computation speed and also improves the parameter search, while retaining a similar ability to fit the data and
forecast accuracy as the Pareto/NBD model. The BG/CNBD model assumes that users without repeated purchases have not churned (defected) yet, independently of their time of inactivity, which may seem counterintuitive. The MBG/CNBD model, employing the Markov--Bernoulli geometric distribution \cite{batislam2007empirical}, eliminates this inconsistency by assuming that users without any activity remain inactive, thus yielding more plausible estimates for the dropout process.

More recent parametric methods include the BG/CNBD-$k$ and MBG/CNBD-$k$ formulations \cite{platzer2016ticking,platzer2016customer}, which extend the BG/NBD and MBG/NBD models, respectively. They consider a fixed regularity within transaction timings, i.e. purchase times are Erlang-$k$ distributed \cite{herniter1971probablistic}. If purchase timings are regular or nearly regular, these models can provide significant improvements in forecasting accuracy without increasing the computational cost.

\subsection{Deep multilayer perceptron}

A deep multilayer perceptron \cite{bengio2009learning} is a type of deep neural network (DNN). In the field of game data science, DNNs have been applied to the prediction of both churn \cite{kim2017churn} and purchases \cite{sifa2018customer}, and also to the simulation of in-game events \cite{guitart2017forecasting}. While \cite{sifa2018customer} focused on predicting the total purchases \emph{over one year} from the player activity within their first seven days in the game, in this work we are interested in estimating the purchases a player will make since the day of the prediction until they leave the game, that is, over a time period that may range from a couple of days to several years. 

A deep multilayer perceptron consists of an input layer, multiple hidden layers, and an output layer \cite{schmidhuber2015deep}. Features (user activity logs) constitute the input of the input layer, and the prediction result (LTV) is the output of the output layer. Layers are connected and are formed by neurons with nonlinear activation functions. Multiple iterations, known as epochs, are performed to optimize the neural network during the learning process. In each epoch, a gradient descent algorithm adjusts weights with the aim of minimizing the value of a predefined cost function, e.g. the root-mean-square error. 

In our analysis, samples were divided into a training and a validation set, used to train and validate the DNN in each epoch, respectively. Moreover, early stopping was applied to prevent overfitting \cite{prechelt1998early}.

\subsection{Convolutional Neural Networks}

A CNN is a type of DNN with one or more convolutional layers that are typically followed by pooling \cite{scherer2010evaluation} and fully connected layers \cite{lecun1989backpropagation, szegedy2015going}. Filters (kernels) are repeatedly applied over inputs in the convolutional layers. With filters that cover more than one input, CNNs can learn local  connectivity
between inputs \cite{turaga2010convolutional}. While deep multilayer perceptrons require feature engineering to transform time-series game logs into structured data (e.g. it is necessary to calculate playtime statistics from the daily playtime time series), CNNs are able to learn user behavior directly from the raw time series.

CNNs have been widely used in image and signal processing, and also for time series 
prediction \cite{lecun1995convolutional}. For instance, in \cite{Babu2016Deep}, the remaining useful life of system components is predicted using CNNs and time series data from sensors. In \cite{tsantekidis2017forecasting}, CNNs were applied to the forecast of stock prices, using as input time series data on millions of financial exchanges. Finally, in \cite{yang2015deep}, human activity recognition was performed by modeling the time series data obtained from body sensors. In this work, we applied CNNs to multichannel time series data from player activity logs to learn player behavior over time and predict future purchases.




\section{Lifetime value using deep learning and parametric models}


\subsection{Model specification}

Two types of DNN models, a deep multilayer perceptron model and a CNN model, were explored.

The deep multilayer perceptron consisted of five fully connected layers. (One input layer, three hidden layers, and one output layer.) In the input layer there were 203 nodes, matching the number of selected features. The hidden layers had 300, 200, and 100 neurons. Weights were initialized by Xavier initialization \cite{glorot2010understanding} and ADAM \cite{kingma2014adam} (a gradient descent optimization algorithm) was used to train the network. The activation functions were sigmoid functions.

\begin{figure*}
  \centering
  \includegraphics[width=\textwidth]{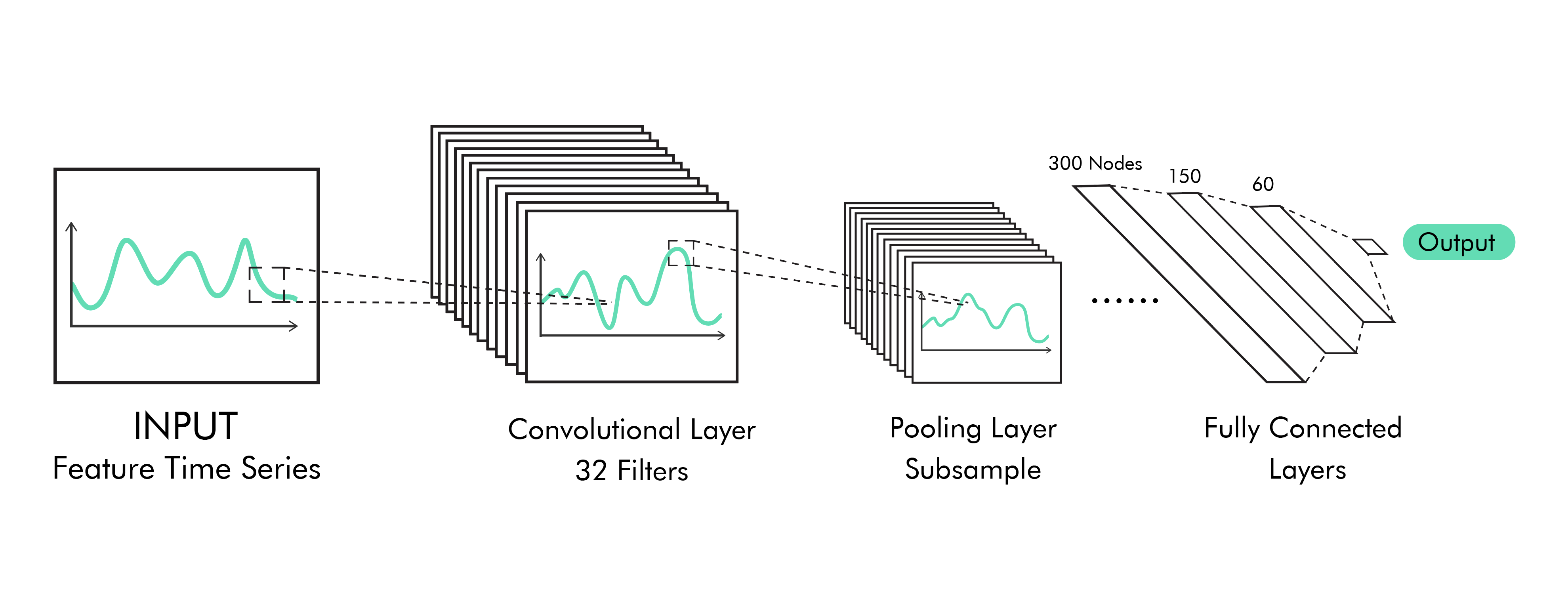} 
\caption{Structure of the convolutional neural network used in this study. It consists of an input layer, convolutional layers, a max pooling layer, fully connected layers, and an output layer. The inputs are time series illustrating player behavior, and the output is the predicted amount of money every player will spend until they leave the game.}
\label{cnn}
\end{figure*}

The CNN model consisted of ten layers. Sequentially, these were an input layer, the first convolutional layer, a max pooling layer, the second convolutional layer, the third convolutional layer, a flatten layer, three fully connected layers, and an output layer. The structure is shown in Figure \ref{cnn}. The number of filters in the three convolutional layers was 32, 16, and 1,
and their size was 7, 3, and 1, respectively. The pool size of the max pooling layer, which controls overfitting \cite{scherer2010evaluation}, was 2. The three fully connected layers had 300, 150, and 60 nodes. Xavier initialization and the ADAM optimization algorithm were also applied. In this case, the activation functions were rectifier functions \cite{glorot2011deep}.

For both deep learning models, data from churned users (those who already left the game) were separated into a training set (80\% of the users, where 20\% of them are used to validate, and a test set (the remaining 20\%). During every epoch, the deep learning models were updated using the training set, and then predictions were performed for the  validation set. Once prediction errors in the validation set did not decrease for 20 epochs, iterations were stopped and the model with the lowest validation error was adopted.\looseness=-1 

The features for the deep learning models consisted of behavior logs for every individual player, including information about game levels, playtime, sessions, actions, and purchases. The features for the CNN model were the daily time series of the logs mentioned above, since the user started playing the game until the day predictions were performed. The features for the deep multilayer perceptron model were the statistics of the logs, such as the average daily playtime or the maximum number of level-ups between two consecutive purchases.

To find the LTV of the players, we need to predict how many purchases they will make until they quit the game. We studied the performance of four different parametric models (Pareto/NBD, BG/NBD, BG/CNBD, MBG/CNBD), all of which require a fixed prediction horizon, which was set to 365 days. Finally, to estimate the value of future purchases, we explored two different approaches: using gamma distributions or simply assigning the average spend per purchase extracted from each player's transaction history. A truncated gamma distribution was also considered, but the results are not shown here, as they were almost identical to the non-truncated case.\looseness=-1




\subsection{Dataset}



The dataset used in the present work comes from \emph{Age of Ishtaria}, a role-playing, freemium, social mobile game with several millions of players worldwide, originally developed by Silicon Studio. Different kinds of in-app purchases, including \emph{gachas} (a monetization technique, very popular in Asian games, that consists on getting an item at random from a large pool of items), are available in this game.

One of the main motivations for this study is finding a suitable method to detect top spenders (who are called \emph{whales} in the game industry) as soon as possible. These players are of the utmost importance because, despite being a small minority (around 2\% of all players), they may provide up to 50\% of the total revenue of the game. 

\begin{figure}
  \centering
  \includegraphics[width=\columnwidth]{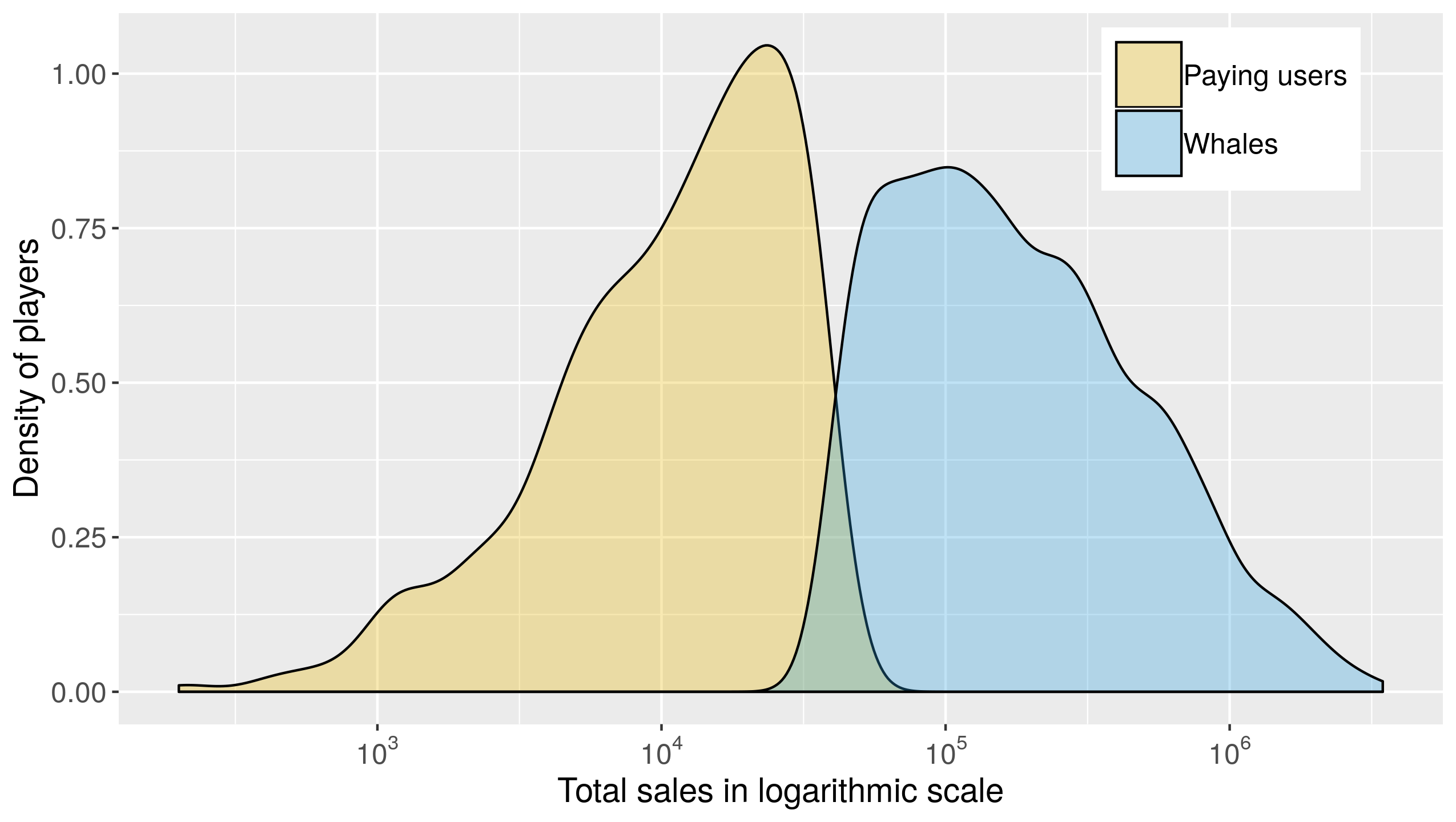} 
\caption{Probability density function from the kernel density estimation of total sales for paying users and \emph{whales} (top spenders).}
\label{distribution}
\end{figure}

We can define whales as those players
whose total expenditure exceeds a certain threshold, which can be computed using the first months of available data as follows: players are sorted by the amount they spent over a given month, and then the threshold is set at the point where the cumulative expenditure reaches 50\% of the total revenue. We repeat this procedure for a number of months and get the final (monthly) threshold as the average of the different values obtained. Figure \ref{distribution} shows the probability distributions of sales (derived from the kernel density estimation) for those top spenders and for the rest of paying users (PUs). The $x$-axis represents total sales in yens (using a logarithmic scale) and the area under each probability density function integrates to 1. We observe there is a meaningful difference between both distributions.

\begin{figure}
  \centering
  \includegraphics[width=\columnwidth]{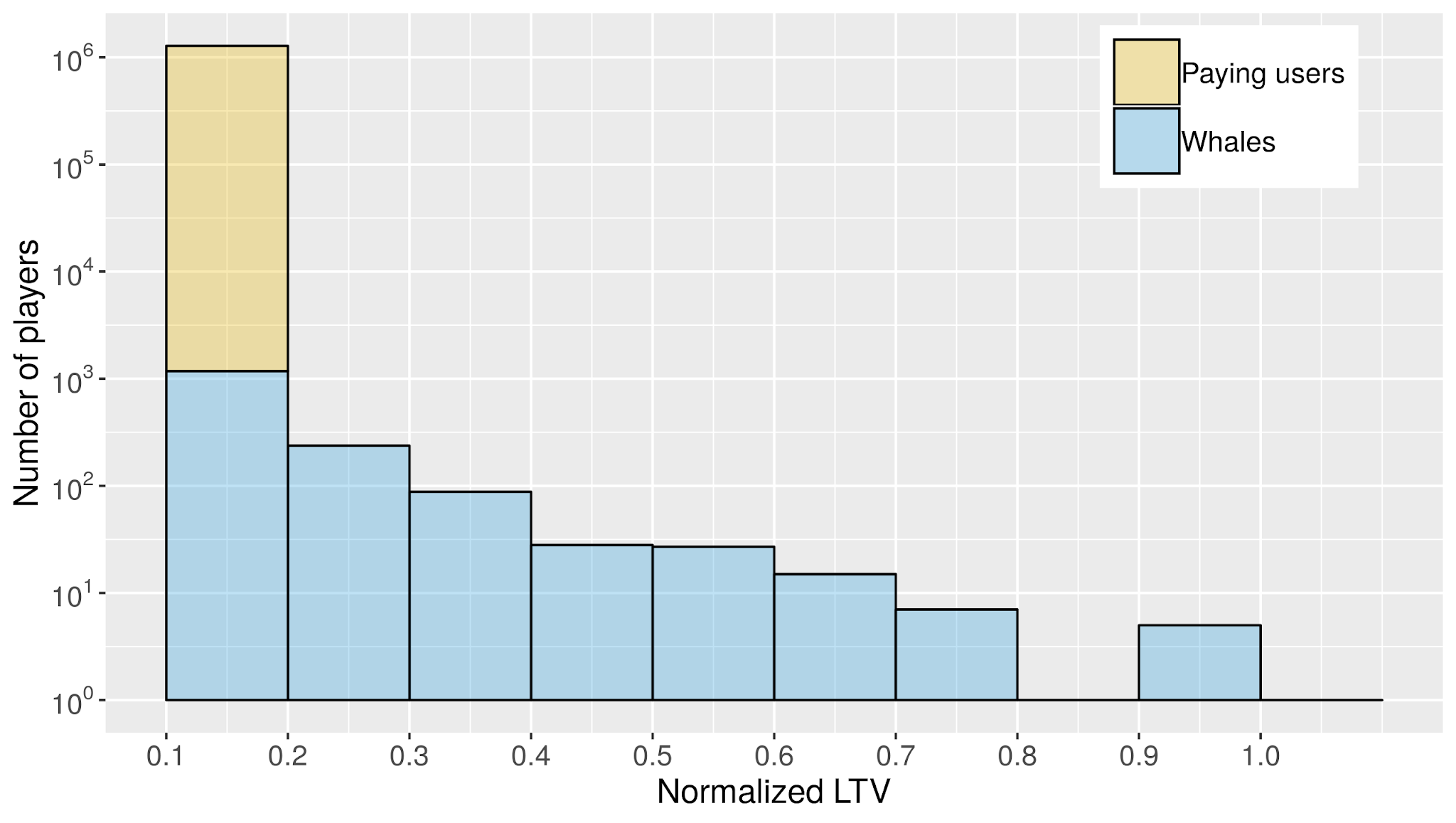}
\caption{Histogram of the number of paying users and whales by lifetime value (normalized between 0 and 1).}
\label{LTV_distribution_PU}
\end{figure}

Figure \ref{LTV_distribution_PU} depicts the distribution of whales and PUs by normalized LTV value. 
As expected, there are few whales with very large LTV values---but these are the most important players in terms of revenue. 

The time period covered by our dataset goes from 2014-09-24 to 2017-05-01, amounting to about 32 months of data. However, a previous history of transactions is needed in BTYD models, and also for feature construction in DNN models. To meet this requirement, we took the simple approach of limiting our study to PUs who were active from 2016-05-01 to 2017-05-01, using the previous data to extract the RFM information and relevant features for the DNN training. There were 2505 paying users who churned in that period. It should be noted that the definition of churn in F2P games is not straightforward. As in \cite{perianez2016churn}, we considered that a player had churned after 9 days of inactivity. After inspection of the data, this seems a reasonable definition, as players who remained inactive for 9 days and then became active again in the future (i.e. players incorrectly identified as churners) contribute marginally to the monthly revenue of the game (far less than 1\%).  


\subsection{Predictor variables}

\begin{figure}[t]
  \centering
  \includegraphics[width=\columnwidth]{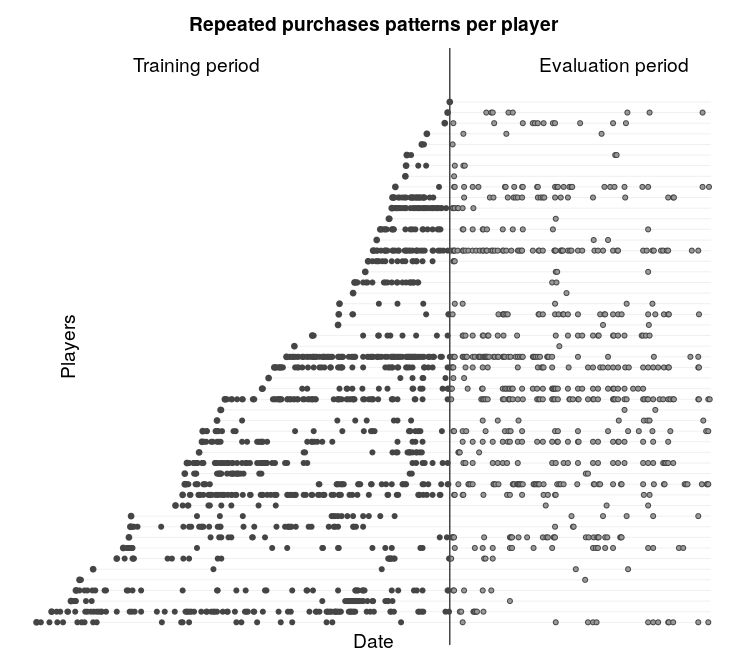} 
\caption{Player purchasing patterns for a sample of paying users during the training and evaluation (prediction) periods.}
\label{purchasePattern}
\end{figure}

\begin{figure}
  \centering
  \includegraphics[width=\columnwidth]{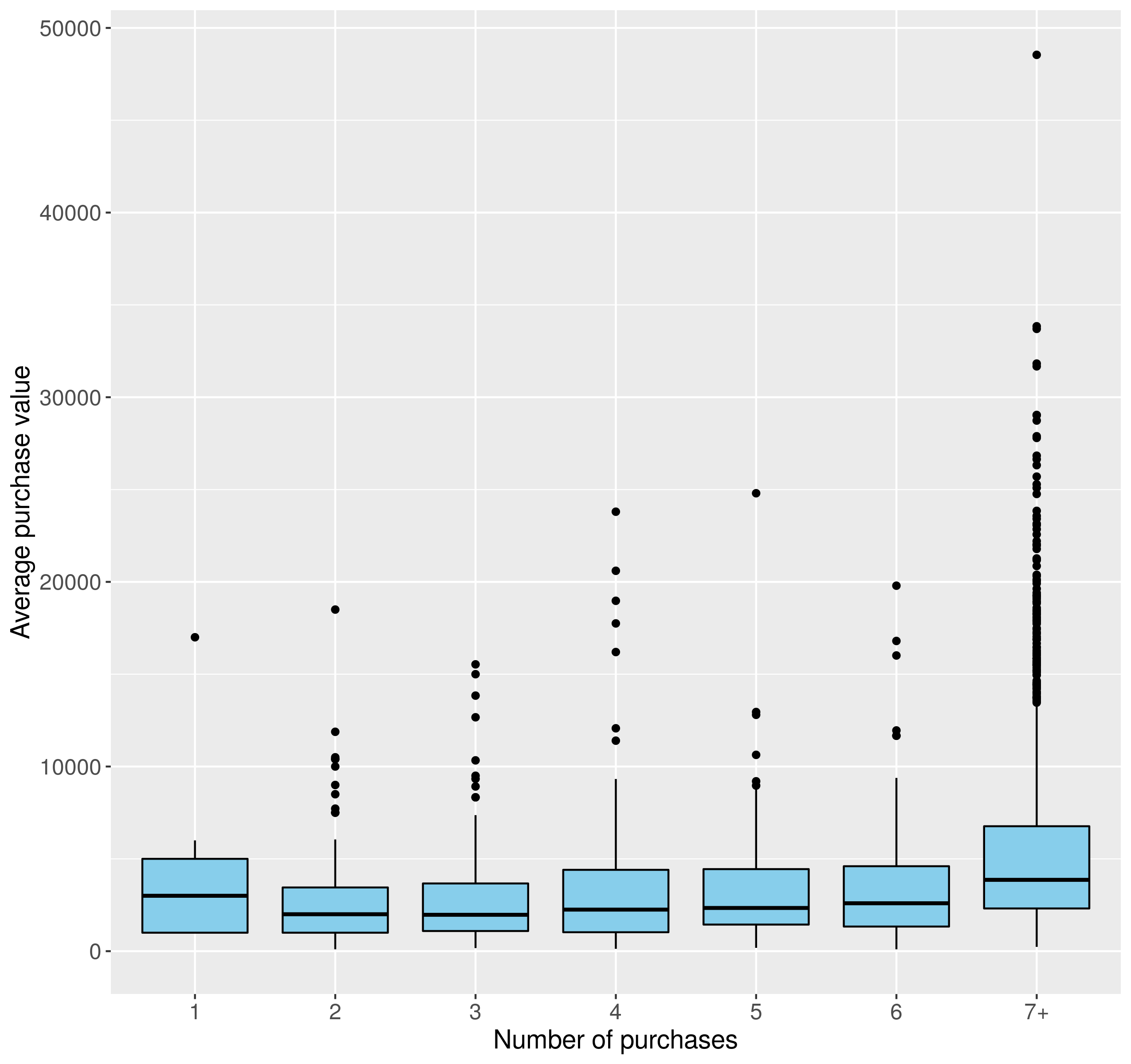}
\caption{Distribution of the average purchase value as a function of the number of repeated purchases for all paying users.}
\label{averagePurchasePerNumPurchases}
\end{figure}

In the case of parametric models, the only information to be considered is the individual purchase information of each player. Figure \ref{purchasePattern} shows the purchasing patterns for a sample of users. The predictor variables for this kind of models (recency, frequency, and monetary value) can be directly extracted from these data. In Figure \ref{averagePurchasePerNumPurchases}, the average purchase value distribution as a function of the number of purchases is represented through a boxplot.

In deep learning models, additional player information can also be taken into account. Feature engineering and data preparation was performed similarly as in \cite{perianez2016churn,GameBigData}. Features were constructed from general data on player behavior that are present in most games (as we want our method to be applicable to various kinds of games, namely to different data distributions), such as daily logins, playtime, purchases, and level-ups.\looseness=-1 

\section{Results}

Although we used four different parametric models to predict LTV, the main differences in performance appeared between those allowing for regularity within transaction timings (i.e. those using CNBDs) and those that consider a gamma distribution for the purchase rate. Therefore, for clarity, we will focus only on two of these methods: the Pareto/NBD model (which is regarded as a benchmark model for these kind of calculations) and the MBG/CNBD-$k$ model (which produced the best results among the four parametric models considered).

The results are summarized in Tables~\ref{BrierTableTraining} and \ref{BrierTablePrediction}, which show four different verification measures for the training and the predictions of the various models. These error estimation metrics are the root-mean-square logarithmic error (RMSLE), the normalized root-mean-square error (NRMSE), the symmetric mean absolute percent error (SMAPE), and the percent error. While the RMSLE is sensitive to outliers and scale-dependent \cite{hyndman2006another}, the NRMSE is more convenient to compare datasets with different scales. The SMAPE is based on
the absolute differences between the observed and predicted values divided by their average.  
Percent errors in this work are calculated as the mean of the deviations divided by the maximum observed value. 

All models produce predictions with percent error (as defined above) below 10\%. Both neural networks outperform all parametric models, significantly improving the accuracy 
with respect to the benchmark Pareto/NBD model and with similar NRMSE values to those found in \cite{sifa2018customer} for high-value players. 
On the other hand, we see that the DNN and CNN models present almost identical results, both for the training and the predicted values. This similarity is probably largely explained by the high overlap among the features used in both models. 

Concerning the parametric models, as already noted above, introducing some complexity (i.e.\ allowing for regularity) in the prediction of the number of future purchases yields improved results. Introducing gamma submodels for the spend per purchase (as opposed to simply taking the average of each player's historic purchases), however, hardly has any impact. This suggests that, in production environments using this type of models, it is probably not justified to invest many resources in introducing more sophisticated models that deal with this issue.

In particular, the significant reduction in RMSLE observed for CNN/DNN models suggests that they perform better than BTYD models at all scales. Indeed, a closer inspection of the data reveals that the parametric models, despite showing a comparable accuracy to deep learning techniques for 
users in a certain range, 
share two problems. First, in many cases they (wrongly) predict no purchases for players who actually keep on spending. The second issue is particularly relevant for the problem at hand: they systematically underestimate the expenditure of top-spending players, i.e. they are particularly ill-suited to describe (and thus to detect) the purchasing behavior of the highest-value players. 

We can readily see this second problem in Table \ref{BrierTablePredictionComp}, which compares the prediction errors for all PUs to those computed for the 20\% of players that spent the most during that year. 
%
While relative errors increase significantly in all models when considering only top spenders, the performance of DNN and CNN models remains much better, with errors that are roughly half as large as in parametric models.

It might come as no surprise that deep learning models outperform the simpler stochastic models: they make use of much more information and, in the case of DNNs, are also considerably more expensive in terms of computational resources. The early detection of high value players, however, is a complicated problem---and one of the utmost importance. The good performance of deep learning methods in this study, even when using a sample of limited size, suggests they have great potential for LTV prediction in production environments, particularly in the case of larger titles (i.e. AAA games, with more paying users and datasets that span longer periods of time).

\begin{table}[tbh!]
	\centering
	\caption{Error measures for the LTV training}
	\begin{tabular}{@{}lcccc@{}} \toprule
Model   & RMSLE & NRMSE & SMAPE & \% Error \\ \midrule
Pareto/NBD + average    & 9.42 & 1.89 & 95.87 & 6.20\% \\ 
Pareto/NBD + gamma      & 9.43 & 1.91 & 96.29 & 6.24\% \\ 
MGB/CNBD-k + average    & 3.41 & 1.72 & 75.44 & 5.52\% \\ 
MGB/CNBD-k + gamma      & 3.55 & 1.77 & 78.58 & 5.71\% \\ 
DNN                     & 1.78 & 1.07 & 75.08 & 3.90\% \\
CNN                     & 1.74 & 1.11 & 72.75 & 3.96\% \\
	\end{tabular}
	\label{BrierTableTraining}
\end{table}

\begin{table}[tbh!]
	\centering
	\caption{Error measures for the LTV prediction}
	\begin{tabular}{@{}lcccc@{}} \toprule
Model     & RMSLE & NRMSE & SMAPE & \% Error \\ \midrule
Pareto/NBD + average      & 9.35 & 1.88 & 95.65 & 8.96\% \\ 
Pareto/NBD + gamma        & 9.37 & 1.88 & 96.35 & 9.01\% \\ 
MGB/CNBD-K + average      & 3.46 & 1.68 & 75.53 & 7.96\% \\ 
MGB/CNBD-K + gamma        & 3.61 & 1.73 & 79.67 & 8.16\% \\ 
DNN                       & 1.82 & 1.12 & 72.99 & 5.82\% \\
CNN                       & 1.84 & 1.05 & 73.76 & 5.72\% \\
    \end{tabular}
	\label{BrierTablePrediction}
\end{table}

\begin{table}[tbh!]
	\centering
	\caption{Prediction error for top spenders}
	\begin{tabular}{@{}lcc@{}} \toprule
Model     & \% Error (All PU) & \% Error (Top Spenders) \\ \midrule
Pareto/NBD + average      & 8.96\% & 33.35\% \\ 
Pareto/NBD + gamma        & 9.01\% & 33.39\% \\ 
MGB/CNBD-K + average      & 7.96\% & 29.14\% \\ 
MGB/CNBD-K + gamma        & 8.16\% & 30.20\% \\ 
DNN                       & 5.82\% & 15.76\% \\
CNN                       & 5.72\% & 15.64\% \\
    \end{tabular}
	\label{BrierTablePredictionComp}
\end{table}




\section{Summary and Conclusion}

Lifetime value is an estimate of the remaining revenue that a user will generate from a certain moment until they leave the game. We profiled players according to their playing behavior and used machine learning to predict their LTV. Deep learning methods were evaluated and compared to parametric models, like the Pareto/NDB model and its extensions. CNN and DNN approaches not only show higher accuracy, but---more 
importantly---such improved performance stems from significantly better predictions for top spenders, whose purchasing behavior is very poorly captured by BTYD models. These users are of paramount importance, as they may generate up to 50\% of all the game revenue, and thus their early detection is one of the primary aims of the player LTV prediction. 

The results were examined not only in terms of accuracy, but also from an operational routine and computational efficiency perspective. The ultimate goal is to find a model that can be run in a production environment on a daily basis and is able to analyze the big datasets generated by players---including their log actions and behavioral records---since they join the game. 

Further work will focus on assessing the sensitivity of our predictions to the forecasting horizon, automatically determining the optimal horizon for each game, and finding the minimum training set size that still yields accurate results. We also plan to extend the evaluation to larger datasets, where we anticipate that the relative gain in accuracy provided by deep learning approaches should become much larger. Moreover, for CNNs we also expect significant savings in computational time, as these networks are able to assimilate raw data---so they do not require data pre-processing or feature engineering, as the deep multilayer perceptron or Pareto/NDB models. In the case of AAA video games, where datasets can be extremely large (easily of the order of petabytes), this advantage could prove essential.

Additionally, we plan to evaluate other deep learning structures, such as long short-term memory (LSTM) networks \cite{Hochreiter1997Long}, which consist of LSTM layers followed by fully connected layers and use time-series records as inputs. While CNNs focus more on modeling timestamp relations, LSTM layers (thanks to their longer memory) can learn feature representations from time-series data with a long-term view. Then, the fully connected layers are able to predict the amount of purchases from those feature representations. 

CNNs emerge as a promising technique to deal with massive amounts of sequential data, a major challenge in video games, without previous manipulation of the player records. For LTV computations, CNNs provide more accurate results, an effect that is expected to increase with the size of the dataset.



\section*{Acknowledgements}
We thank Javier Grande for his careful review of the manuscript and Vitor Santos for his support.


\bibliographystyle{abbrv}
\bibliography{main.bib}	

\end{document}